\documentclass[
 reprint,
 amsmath,amssymb,
 aps,
 prl,
]{revtex4-2}

\usepackage{graphicx}% Include figure files
\usepackage{dcolumn}% Align table columns on decimal point
\usepackage{bm}% bold math
\usepackage{multirow}
\usepackage{subfigure}
\usepackage{textcomp}
\usepackage[T1]{fontenc}
\usepackage[utf8]{inputenc}
\usepackage{mathptmx}
\usepackage[svgnames]{xcolor}

\newcommand{\resubmitedit}[1]{{\textcolor{Black}{{#1}}}}

\begin{document}

% \title{\resubmitedit{Molecular Dynamics Perspectives on Multi-Scale Modeling of Non-Ideal Fluids}}

\title{Molecular dynamics perspectives on nonideal fluid models for the lattice Boltzmann method}

\author{Hiroshi Otomo}
\email{hiroshi.otomo@tufts.edu}
\affiliation{Department of Mathematics, Tufts University, Medford, Massachusetts 02155, USA}

\author{Alexander J. Wagner}
\affiliation{Department of Physics, North Dakota State University, Fargo, North Dakota 58108, USA}

\begin{abstract} 
Despite their widespread use, mesoscopic models for non-ideal fluids have rarely been systematically validated against microscopic simulations. In this work, molecular dynamics (MD) simulations of confined fluids are mapped onto a mesoscopic framework, enabling direct comparison with lattice Boltzmann (LBM) formulations. 
By analyzing the moments of the distribution function, we identify a force formulation that consistently reproduces the microscopic statistics and macroscopic force balance. The results show that a hybrid formulation combining pseudo-potential and free-energy approaches provides the most consistent description. 
These findings establish a direct link between microscopic particle dynamics and mesoscopic modeling, offering practical guidance for the development and selection of LBM models for non-ideal and multiphase flows.  \footnote{Accepted for publication in Physical Review Fluids. The definitive peer-reviewed version is available at  \url{https://aps.org}}
\end{abstract}

\maketitle
 %%%%%%%%%%%%%%%%%%%%%%%%%%%%%%%%%%%%%%%%%%%%%%%%%%%%
 %%%%%%%%%%%%%%%%%%%%%%%%%%%%%%%%%%%%%%%%%%%%%%%%%%%%
  %%%%%%%%%%%%%%%%%%%%%%%%%%%%%%%%%%%%%%%%%%%%%%%%%%%%

\section{Introduction} % 272->278 words (including the citation sentence)
\label{Intro}

The gap between microscopic and macroscopic fluid dynamics remains a central challenge due to computational and modeling limitations in fields such as computational fluid dynamics and fluid engineering \cite{Succi2018,Schiller2018}. In particular, the behavior of non-ideal fluids involves complex particle interactions and thermodynamic effects, making multiscale modeling essential.
To bridge microscopic and macroscopic descriptions, several mesoscopic approaches—including the lattice Boltzmann method (LBM) \cite{2010_review,kruger2017lattice,OTOMO2023}, dissipative particle dynamics \cite{Groot_1997,Jamali_2015}, and multi-particle collision dynamics \cite{Zantop_2021} —have been developed over the past decades to capture non-ideal fluid behavior. Among these, LBM has become a widely used tool for simulating non-ideal and multiphase flows due to its computational efficiency and flexibility in modeling interfacial phenomena.
Despite their practical success, mesoscopic models are often constructed with limited direct validation against microscopic simulations such as molecular dynamics (MD) \cite{Denniston_2004,Tong_2022,Groot_1997,Jamali_2015,Zantop_2021}. In particular, multiple force formulations within the LBM framework—such as pseudo-potential and free-energy approaches—have been proposed and developed independently, and systematic criteria for selecting or relating these models to underlying microscopic physics remain unclear \cite{Gunstensen_1991,Shan-Chen_1993,Swift_1995,Swift_1996,Guo_2002,EDM_2009}.
In LBM, the evolution of a discrete distribution function provides a statistical representation of particle dynamics and serves as a key mesoscopic quantity connecting microscopic behavior to macroscopic flow properties. Establishing a physically grounded multiscale framework therefore requires examining the consistency of this quantity between microscopic and mesoscopic descriptions, including its implications for thermodynamic consistency and force balance.
In the present study, MD simulations of particles confined by reflective walls under external forcing are employed to provide a microscopic reference for mesoscopic modeling. Using a systematic mapping from MD to the LBM~\cite{Alex_2017}, we analyze how the statistical properties of particle distributions relate to force formulations in the lattice Boltzmann models. This approach enables a direct comparison of distribution function moments and provides insight into how non-ideal interactions and thermodynamic effects are represented at the mesoscopic level. The main findings are summarized in the concluding section.

 %%%%%%%%%%%%%%%%%%%%%%%%%%%%%%%%%%%%%%%%%%%%%%%%%%%%
 %%%%%%%%%%%%%%%%%%%%%%%%%%%%%%%%%%%%%%%%%%%%%%%%%%%%

\section{Molecular dynamics simulation} %553 words + 1eq(16*1=16) + 1figure(150*0.7+20)+caption(19) =713
\label{sec_MDsimulation}
The motion of gas particles under a spatially uniform external force is simulated using LAMMPS~\cite{LAMMPS,plimpton1995fast}.  
For interparticle interactions, we employ the Lennard-Jones (LJ) potential,  
$V(r) = 4 \epsilon_{LJ} \left[ \left(\sigma/r \right)^{12} - \left( \sigma/r \right)^6 \right]$  for $ r \le r_c$,
and \( V(r) = 0 \) for \( r > r_c \), with cutoff \( r_c = 5\sigma \), which is sufficient for the present low-density conditions.  
Here, \( \epsilon_{LJ} \) is the potential well depth and \( \sigma \) the particle diameter.
Throughout this study, all quantities are non-dimensionalized: lengths by $\sigma$, energies by $\epsilon_{LJ}$, and time by $\sqrt{m\sigma^2/\epsilon_{LJ}}$, where $m$ is the particle mass.
A total of 12,\!000 particles are placed in the simulation domain $\Omega \in [-15,15] \times L_{yz}$, where the spanwise domain size is $L_{yz} = 200 \times 200$. This configuration provides sufficient sampling of particles and results in an average number density of $\rho = 0.01$. Reflective walls are imposed at the $x$-boundaries, while periodic boundary conditions are applied in the $y$- and $z$-directions.
Particle trajectories are computed by integrating Newton's equations of motion using the velocity-Verlet scheme, coupled with a Nosé–Hoover thermostat. The temperature is fixed at $T = 1.0$, and the time step size is set to $0.001$.
Simulations are run for $6.0 \times 10^6$ time steps.
All reported statistical quantities are obtained by averaging over the last $4.0 \times 10^6$ time steps and within measurement boxes of size $L_x \times L_{yz}$ at different x positions, where $L_x = 1$, after confirming that the system has reached a statistical steady state. A spatially uniform external force is applied in the $x$-direction, with a sufficiently large acceleration $g_x = 0.1$ or $g_x = 0.3$ to realize non-ideal gas behavior. As a result, from a statistical perspective, the system can be regarded as quasi-one-dimensional.

\begin{figure}[t]
\centering
\includegraphics[width=0.95\linewidth]{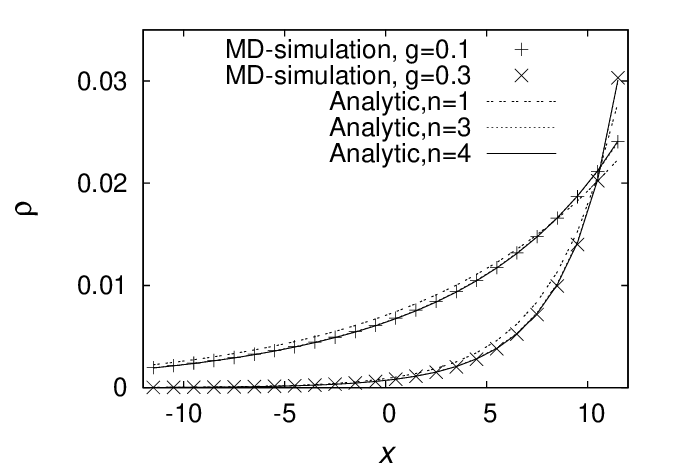}
\caption{Number density profiles along $x$ for $g=0.1$ and $g=0.3$, with analytic solutions at first-, third-, and fourth-order in $\epsilon$.}
\label{fig_density_all}
\end{figure}

The averaged simulation results over each time window of $1.0 \times 10^6$ time steps yield static density profiles, with the macroscopic flow velocity remaining negligibly small (less than $10^{-5}$).  
The resulting density profiles $\rho(x)$ for each value of $g_x$ are shown in Fig.~\ref{fig_density_all}, where the $x$-range is chosen to minimize boundary effects.
According to fluid dynamics theory, the steady-state density profile satisfies the force balance equation, $dP_{\mathrm{EOS}}/dx = \rho g_x$,
%\begin{align}
%\label{force_balance}
%\frac{dP_{\mathrm{EOS}}}{dx} = \rho g_x,
%\end{align}
where $P_{\mathrm{EOS}}$ denotes the pressure as a function of $\rho$ defined by an equation of state (EOS).
To account for deviations from ideal gas behavior under compression, we adopt a cubic EOS derived from the modified Benedict–Webb–Rubin (MBWR) equation~\cite{Teja_1996}, $P_{\mathrm{EOS}} = \sum_{k=1}^{3}  \kappa_k \rho^k$ with coefficients $\kappa_1 = 1.0$, $\kappa_2 = -5.291$, and $\kappa_3 = 4.695$ at $T = 1$.
Using this equation of state, the force balance equation is approximately solved by assuming a perturbative expansion of the form
\begin{align}
\label{rho_anal}
\rho_{\mathrm{anal}}(x) = 
 \sum^{n}_{k=1} \epsilon^k \rho_k  \exp(k g_x x),
\end{align}
where $n$ denotes the expansion order, and the solution is organized in powers of $\epsilon$.
The coefficients up to $n=4$ are
$\rho_1 = 1$,
$\rho_2 = -2 \kappa_2$,
$\rho_3 = -\frac{3}{2}(\kappa_1 g - 4 \kappa_2^2)$, and
$\rho_4 = 4 \kappa_2 \left( 2 \kappa_1 +  \kappa_1  g - 4 \kappa_2^2 - \frac{4}{3} \kappa_2 \right).$
As shown in Fig.~\ref{fig_density_all}, the series converges well with respect to $\epsilon$, and agreement with the simulation results improves as higher-order terms are included. Specifically, the maximum deviation from the simulation is $0.77\%$ for $g_x = 0.1$ with expansion up to third order using $\epsilon = 6.01 \times 10^{-3}$, and $1.32\%$ for $g_x = 0.3$ with expansion up to fourth order using $\epsilon = 7.07 \times 10^{-4}$.

 %%%%%%%%%%%%%%%%%%%%%%%%%%%%%%%%%%%%%%%%%%%%%%%%%%%%
 %%%%%%%%%%%%%%%%%%%%%%%%%%%%%%%%%%%%%%%%%%%%%%%%%%%%
 
\section{Non-ideal fluids model of the lattice Boltzmann method (LBM)} 
%567 words +9eq(9*16=160) + 1table (13+6.5*4=39) = 750
\label{LBM_force_scheme}

In the lattice Boltzmann method (LBM), the Boltzmann equation is solved in discrete space and time. 
Particle velocities are represented by a finite set $\{c_i\}$, where each index $i$ denotes both the direction and magnitude of a discrete velocity. 
This discretization naturally introduces spatial and temporal resolutions, $\Delta x$ and $\Delta t$, respectively. 
The particle distribution function for velocity $c_i$ at position $x$ and time $t$ is denoted by $f_i(x,t)$.
The discrete Boltzmann equation in this framework takes the standard LBM form
\begin{align}
\label{Org_LB}
f_i \left( x + c_i \Delta t , t + \Delta t \right) - f_i \left( x , t \right)
   = \Omega_i \left( x, t \right) + F_i \left( x, t \right),
\end{align}
where $\Omega_i$ is the collision operator and $F_i$ represents the forcing term. 
For the single-relaxation-time Bhatnagar-Gross-Krook (BGK) model, these are expressed as
\begin{align}
\label{col_term}
\Omega_i = - \frac{f_i - f^{\mathrm{eq}}_i}{\tau}, \qquad
F_i = w_i \gamma_1 \left( \frac{c_{i,\alpha} F_{\alpha}}{c_s^2} 
+ \mathcal{R}_{\mathrm{sec},i} \right),
\end{align}
where $\alpha \in \{x,y,z\}$, $\tau$ is the relaxation time, and $c_s$ is the lattice sound speed.
The coefficient $\gamma_1$ and the term $\mathcal{R}_{\mathrm{sec},i}$, 
which satisfies $\sum_i w_i \mathcal{R}_{\mathrm{sec},i} c_{i,\alpha}^k = 0$ for $k \le 1$, 
are determined by the choice of force scheme.
Macroscopic quantities such as the density $\rho$ and the fluid velocity $u_F$ are computed from the distribution function $f_i$ as
\begin{align}
\label{rho_u_form}
\rho = \sum_i f_i, \hspace{0.2in} u_{F,\alpha} = u_{\alpha} + \frac{F_{\alpha}}{2\rho},
\end{align}
where $u_{\alpha} = \sum_i f_i c_{i,\alpha}/ \rho$.
The prefactor $1/2$ accounts for force averaging before and after collision.
The equilibrium distribution $f_i^{\mathrm{eq}}$ is constructed to satisfy the following moment conditions:
\begin{align}
\label{feq_mom}
\sum_i f_i^{\mathrm{eq}} = \rho, \hspace{0.2in}
\sum_i f_i^{\mathrm{eq}} c_{i,\alpha} = \rho u_{\alpha} + \gamma_2 F_{\alpha},
\end{align}
and
\begin{align}
\label{feq_sec_mom}
\sum_i f_i^{\mathrm{eq}} c^2_{i}= \rho c_s^2 + \left(1 - \theta \right) \delta P
+ \rho \left( u_{\alpha} + \gamma_2 \frac{F_{\alpha}}{\rho} \right)^2.
\end{align}
These lattice Boltzmann (LB) formulations are presented in a unified and comprehensive manner.  
Here, $\delta P$ denotes the pressure contribution arising from deviations from ideal-gas behavior, $\delta P = P_{EOS} - \rho c_s^2$, as modeled by the free-energy approach~\cite{Swift_1996}.  
The force term $F$, in contrast, can be interpreted as the pressure force due to non-ideal interactions, together with any external forcing, in the context of the pseudo-potential model~\cite{Shan-Chen_1993,Guo_2002,EDM_2009},  
\begin{align}
\label{Fx_form}
F_{x}= \rho g_{x} - \theta \frac{\partial \delta P}{\partial x}.
\end{align}
Here, $\theta$ is defined as the switching parameter between the two models: when $\theta = 0$, the purely free-energy model is used, and when $\theta = 1$, the purely pseudo-potential model is used.  
The coefficients $\gamma_1$ and $\gamma_2$ are determined by the force scheme.  
Table~\ref{table_forceschemeparam} summarizes these parameters' formulation for common schemes, including the Shan-Chen ($S$–$C$) model~\cite{Shan-Chen_1993}, Guo's method ($Guo$)~\cite{Guo_2002}, and the exact difference method ($EDM$)~\cite{EDM_2009}.  
In all cases, the resulting macroscopic equations recover the Navier--Stokes equation with the applied force $F$ at leading order.  
\begin{table}[h]
\setlength{\tabcolsep}{12pt} % 
\centering
\caption{Parameters for typical three force schemes \cite{Shan-Chen_1993,Guo_2002,EDM_2009}}
\label{table_forceschemeparam}
\begin{tabular}{lclclc}
\hline
Force scheme & $\gamma_1$ & $\gamma_2$ & $\gamma_3$ \\ \hline
$S\!-\!C$    & $0.0$ & $\tau$ & $1/4$ \\ 
$Guo$        & $1-1/(2\tau)$ & $0.5$ & $0$ \\ 
$EDM$        & $1.0$ & $0.0$ & $\tau^2-\tau+1/4$ \\ \hline
\end{tabular}
\end{table}
Applying the Chapman-Enskog expansion\cite{OTOMO20171000} to Eq.~(\ref{Org_LB}) and taking the first moment under $u_F = 0$ yields:
\begin{align}
\label{balance_force_2}
\sum_i \left( \Omega_i + F_i \right) c_{i, \alpha} = \frac{\partial}{\partial x_{\beta}} \sum_i c_{i,\alpha} c_{i,\beta} f^{eq}_i +  \mathcal{O} \left( \frac{\partial^2 f_i}{\partial x^2} \right).
\end{align}
Using Eq.~(\ref{col_term})-(\ref{feq_sec_mom}) and Table~\ref{table_forceschemeparam}, one obtains $\sum_i \left( \Omega_i + F_i \right) c_{i, \alpha}  = F_{\alpha}$, leading to the force balance equation:
\begin{align}
\label{balance_force}
F_{x} \approx \frac{\partial}{\partial x} \left( \rho c_s^2 + \left(1 - \theta \right) \delta P \right),
\end{align}
in the $x$-direction.
This relation holds for all force schemes in Table~\ref{table_forceschemeparam}.  
Eq.~(\ref{balance_force}) with Eq.~(\ref{Fx_form}) shows that the pressure force from deviations from ideal-gas behavior can be interpreted as either the body force $F_{x}$ or corrections to the pressure tensor from the second moment of $f^{\mathrm{eq}}_i$.
Hence, the free-energy model ($\theta=0$) and the pseudo-potential model ($\theta=1$) yield the same macroscopic equation at leading order, though their mesoscopic mechanisms differ.  
Differences among the models and schemes in Table~\ref{table_forceschemeparam} appear in the higher-order truncation errors of Eq.~(\ref{balance_force}).
For the force scheme, they stem from corrections to the right-hand side of Eq.~(\ref{balance_force_2}), mainly associated with the second moment of the equilibrium state and the force term,
\begin{align}
\label{gam3_def}
\sum_i \left( f_i^{\mathrm{eq}} + \tau F_i \right) c_{i,x} c_{i,x}
 - \left( \rho c_s^2 + (1 - \theta)\delta P \right) %\nonumber \\
% = \rho \left( \gamma_2 - \tfrac{1}{2} \right) g_x^2
% + \tau \gamma_1 \sum_i w_i \mathcal{R}_{\mathrm{sec},i} 
 \equiv \gamma_3 \rho g_x^2 ,
\end{align}
where $u_F = 0$, and the dependence of $\gamma_3$ on the force scheme is summarized in Table~\ref{table_forceschemeparam}.  
This term affects key properties such as viscosity independence and thermodynamic consistency~\cite{Alex_2006,Shan_2008,Baixin_2015}.

%
%%%%%%%%%%%%%%%%%%%%%%%%%%%%%%%%%%%%%%%%%
\section{Molecular dynamics Lattice Boltzmann (MDLB)}  
%%Aspect ratio 0.7,  0.7, 0.7
% Total 781 + 12eq (16*12) + 4 figures ((150*0.7+20)*4)+ caption(120)=1593
\label{sec-LBM}

\begin{figure}[t]
\centering
\includegraphics[width=0.95\linewidth]{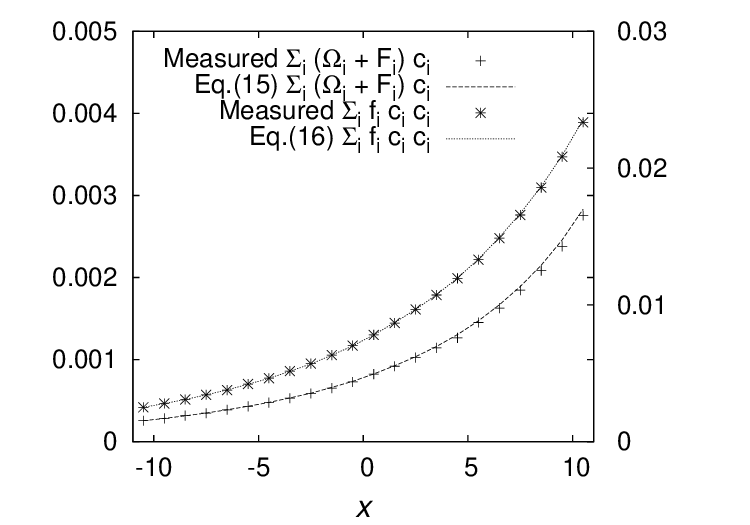}
  \caption{Comparison of the computed $\sum_i (\Omega_i + F_i) c_{i,x}$ with the right-hand side of Eq.~(\ref{fst_omega_ansatz}) (left axis), and of $\sum_i f_i c_{i,x} c_{i,x}$ with the right-hand side of Eq.~(\ref{sec_f_ansatz}) (right axis) for $g=0.1$.}
  \label{fig_Fsec_Omegafst}
\label{fig_mom_devper}
\end{figure}

The MDLB approach \cite{Alex_2017,Alex_2019,Alex_2020_PRE,Alex_2020,Alex_2021,Alex_2021_Phil} interprets MD simulation results within the LBM framework by discretizing space and time with resolutions $\Delta x$ and $\Delta t$ for post-processing.  
The distribution function $f_i$ at $x + \frac{\Delta x}{2}$ and time $t$ is computed as
\begin{align}
\label{def_f}
f_i\!\left( x + \frac{\Delta x}{2}, t \right) = \sum_j \left\langle \Delta_{x}  \!\left[ x_j (t) \right]   \Delta_{x- c_i}  \!\left[ x_j ( t - \Delta t ) \right] \right\rangle ,
\end{align}
where $x_j$ is the $j$-th particle position and
\begin{align}
\Delta_x [x'] = 
\begin{cases}
 1,& \text{if }  x_k < x'_k \le x_k + \Delta x_k,\quad  k \in \{x, y, z\}, \\[2pt]
 0, & \text{otherwise},
\end{cases}
\end{align}
with $x \in |\Delta x| \mathbb{Z}^3$, $x' \in \mathbb{R}^3$, and $\langle \cdot \rangle$ denoting a temporal and cell-volume average.  
The space and time resolution is characterized by
\begin{align}
\label{asquare}
a^2 = \frac{\langle (\delta x)^2 \rangle}{(\Delta x)^2},
\end{align}
where $\langle (\delta x)^2 \rangle$ is the mean-squared displacement over $\Delta t$.  
Here, for simplicity, we set $\Delta x = \Delta t = 1$, yielding $a^2 \in [0.92,\,1.03]$ across the domain (see Supplementary Material).
Assuming $f_i$ in Eq.~(\ref{def_f}) satisfies Eq.~(\ref{Org_LB}) with Eqs.~(\ref{col_term})--(\ref{feq_mom}), we analyze the combined term $\Omega_i + F_i$.  
From Eqs.~(\ref{col_term})--(\ref{feq_mom}),  its first moment becomes $-2 \sum_i f_i c_{i,\alpha}$
with $u_F = 0$.  
Indeed, $\sum_i (\Omega_i + F_i) c_{i,x}$ computed using Eq.~(\ref{Org_LB}) matches $-2 \sum_i f_i c_{i,x}$ for $g = 0.1$ and $0.3$, within $1.9\%$ in $x > 0$ (see Supplementary Material). 
Next, referring Eq.~(\ref{col_term}), the second moment of $\Omega_i + F_i$ is estimated.
Direct evaluation of $\sum_i (\Omega_i + F_i) c_{i,x} c_{i,x}$ from Eq.~(\ref{Org_LB}) and 
$\sum_i f_i c_{i,x} c_{i,x}$  gives characteristic magnitudes of 
$10^{-6}$ and $10^{-2}$, respectively.  
Furthermore, the typical scales of $\rho T$, $\delta P$, and $\rho g_x^2$, arising from 
$f_i^{\mathrm{eq}}$ and $\sum_i w_i \mathcal{R}_{\mathrm{sec},i} c_{i,x} c_{i,x}$, lie in the range $10^{-2}$--$10^{-4}$.  
These results suggest
\begin{align}
\label{f_approx_form}
\sum_i f_i  c_{i,x} c_{i,x} 
   \approx 
   \sum_i \left( f^{\mathrm{eq}}_i  
   + \tau w_i \gamma_1 
   \mathcal{R}_{\mathrm{sec},i} \right) 
   c_{i,x} c_{i,x},
\end{align}
provided $|\tau| \le \mathcal{O}(1)$.
Using this approximation and referring Eqs.~(\ref{Fx_form})--(\ref{gam3_def}), we assume
\begin{align}
\label{fst_omega_ansatz}
\sum_i \left( \Omega_i + F_i \right) c_{i, x} = \Psi_{xx} \left( \Delta t \right)^2 \left( \rho g_{x} - \theta \frac{\partial \left( P_{EOS} - \rho T \right)}{\partial x} \right),
\end{align}
\begin{align}
\label{sec_f_ansatz}
\sum_i f_i c_{i, x} c_{i, x}
= \Psi_{xx} \left( \Delta t \right)^2 \bigg[
    &\left\{ \rho T + \left( 1 - \theta \right)
    \left( P_{\mathrm{EOS}} - \rho T \right) \right\}  \nonumber \\
    &+ \gamma_3 \rho g^2_{x} \left( \Delta t \right)^2
\bigg],
\end{align}
with $u_F=0$.
Here, $\Psi_{xx}(a^2)$ denotes the ratio of the second moment of the discrete MDLB distribution in free space (no external forcing) to its continuum counterpart.  
Originally introduced in Ref.~\cite{Alex_2017}, $\Psi_{xx}$ quantifies discretization effects from the Gaussian displacement probability and links the LBM and MDLB frameworks (see Supplementary Material for the explicit formula).  
In particular, it yields the relation $c_s^2 = \Psi_{xx} \, (\Delta t/ \Delta x)^2 T$ for the steady ideal gas.
For the current setup, $\Delta x = \Delta t =1$, $c_s^2 \approx 1$.
To assess these ansätze, we compare $\sum_i (\Omega_i + F_i) c_{i,x}$ with the right-hand side of Eq.~(\ref{fst_omega_ansatz}) in Fig.~\ref{fig_Fsec_Omegafst} for $g=0.1$.  
The figure also shows a comparison between the measured $\sum_i f_i c_{i,x} c_{i,x}$ and the right-hand side of Eq.~(\ref{sec_f_ansatz}) (results for $g=0.3$ are in the Supplementary Material).  
Here, we set $\theta = 0.5$ and $\gamma_3 = 0.25$.  
They agree very well, with maximum deviations of 3.3\% and 1.0\%, respectively.
Similar levels of agreement are observed for different discretization setups, such as $\Delta x = 1$ and $\Delta t = 0.41$, which yield $c_s^2 \approx 1/ 3$ (see the Supplementary Material).
Figure~\ref{fig_mom_devper} summarizes maximum deviations of $\sum_i (\Omega_i + F_i) c_{i,x}$ and $\sum_i f_i c_{i,x} c_{i,x}$ for different $\theta$ with $\gamma_3=0.25$, evaluated using the $L_{1/2}$ norm to emphasize sensitivity to large deviations atop the geometric mean for $g=0.1$.  
It also shows corresponding deviations of $\sum_i f_i c_{i,x} c_{i,x}$ for different $\gamma_3$ with $\theta=0.5$ at $g=0.3$.  
These results confirm $\theta = 0.5$ and $\gamma_3 = 0.25$ as the most consistent choice.  
The choice $\theta = 0.5$ agrees with a hybrid LB model combining pseudo-potential and free-energy schemes.  
The value $\gamma_3 = 0.25$ rules out a pure Guo scheme and instead indicates S--C, EDM with specific $\tau$, or a hybrid forcing.  
Such higher-order effects are essential for thermodynamic consistency \cite{Alex_2006,Shan_2008,Baixin_2015}, with detailed analysis left for future work.  
\begin{figure}[t]
\centering
\includegraphics[width=0.95\linewidth]{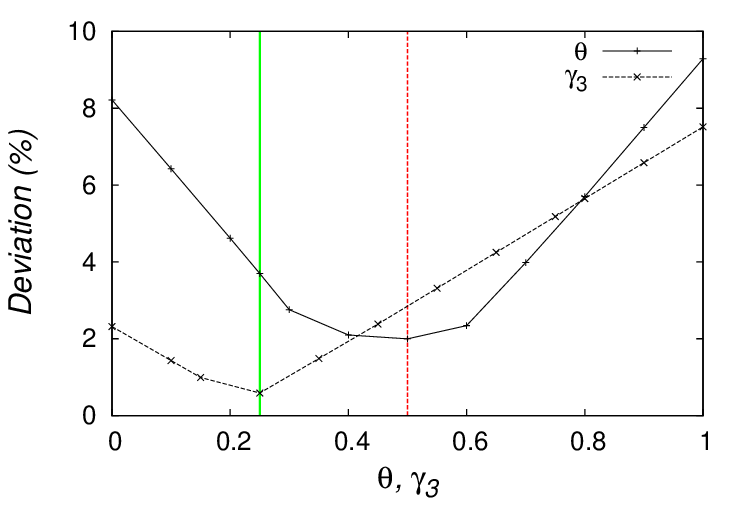}
\caption{Maximum deviation of $\sum_i (\Omega_i+F_i)c_{i,x}$ and $\sum_i f_i c_{i,x}^2$ vs.~$\theta$ with $\gamma_3=0.25$, $g=0.1$ (solid), and of $\sum_i f_i c_{i,x}^2$ vs.~$\gamma_3$ with $\theta=0.5$, $g=0.3$ (dotted). Vertical dotted and solid lines indicate $x=0.5$ and $x=0.25$, respectively.}
\label{fig_mom_devper}
\end{figure}

\begin{figure}[t]
\centering
\includegraphics[width=0.95\linewidth]{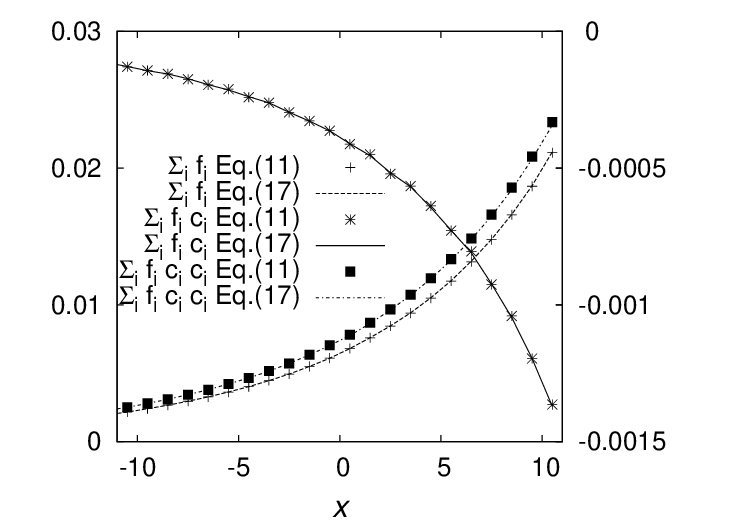}
  \caption{Moments of $f_i$ from Eqs.~(\ref{def_f}) and (\ref{def_f_pbase}) with Gaussian $P(x,\delta x)$ for $g=0.1$. The left axis shows zeroth and second moments, the right axis the first moment.
}
 \label{fig_mom_compared}
\end{figure}

\begin{figure}[t]
\centering
\includegraphics[width=0.95\linewidth]{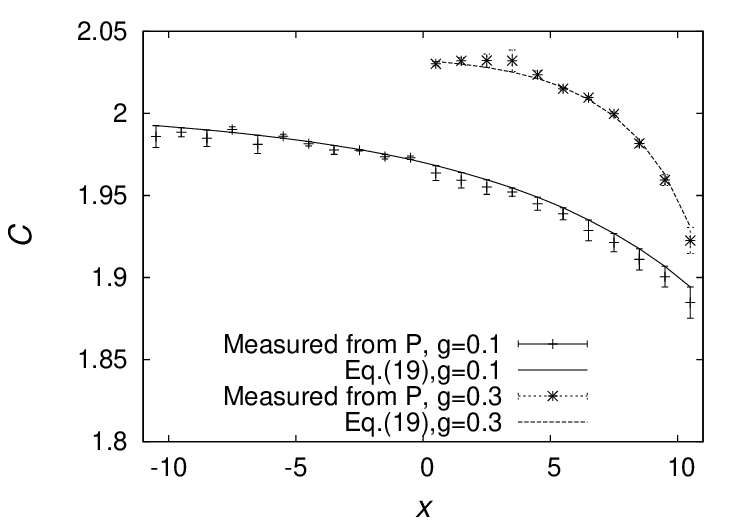}
\caption{Comparison of $C(x)$ from Eq.~(\ref{form_C})  with values measured directly from $P(x,\delta x)$ for $g=0.1$ and $g=0.3$. For $g=0.3$, only sufficiently sampled data at $x>0$ are shown.}
\label{fig_ParamC}
\end{figure}

%Probability of displacement%%%%%%%%%%%%%%%%%%%%%%%%%%%%%%%%%%%%
The displacement probability $P(x,\delta x)$ gives the likelihood that a particle at $x$ and time $t$ was at $x-\delta x$ at $t-\Delta t$, linking continuous dynamics to the discrete representation of $f_i$.  
Using $P$, Eq.~(\ref{def_f}) becomes
\begin{align}
\label{def_f_pbase}
f_i \left( x + \frac{\Delta x}{2}, t \right) = \nonumber \\
\frac{1}{\left( \Delta x \right)^d} \int d \tilde{x} \int d \delta x \, \Delta_x \left[ \tilde{x} \right] \Delta_{x - c_i}\left[ \tilde{x} - \delta x \right] \, \rho(\tilde{x}) P(\tilde{x}, \delta x),
\end{align}
where $d$ is the spatial dimension, generalizing previous work~\cite{Alex_2017} by explicitly including spatial dependence in both $\rho$ and $P$.  
Using the Gaussian $P(x,\delta x) = (1/\sqrt{\pi C}) \exp[-(\delta x - B)^2 / C]$, the double integral is evaluated numerically (Simpson’s rule with linear interpolation for $P$ and $\rho$ from Eq.~(\ref{rho_anal})).  
Figure~\ref{fig_mom_compared} compares results of moments from Eqs.~(\ref{def_f}) and (\ref{def_f_pbase})   for $g = 0.1$ using nine discrete velocity states  showing good convergence.  
The results exhibit excellent agreement, with maximum deviations below 1.4\% across all moments and spatial positions.  
Similar agreement is observed for $g = 0.3$, with deviations within 2.2\% (see Supplementary Material).  
To assess effects from spatial dependence of $\rho$ and $P$, $f_i$ from Eq.~(\ref{def_f_pbase}) is computed with zeroth-order approximations
\[
\rho(\tilde{x}) = \rho\!\left(x + \tfrac{\Delta x}{2}\right), \quad
P(\tilde{x},\delta x) = P\!\left(x + \tfrac{\Delta x}{2}, \delta x\right),
\]
and compared with Eq.~(\ref{def_f}) for $g=0.1$ and $g=0.3$.  
Second moments agree within 1.4--2.2\%, while first moments deviate up to 25\%, indicating that the zeroth-order approximation suffices for practical use of the second moment.  
Following Parsa and Wagner~\cite{Alex_2017}, this yields
\begin{align}
\sum_i f_i \left(x + \frac{\Delta x}{2}, t\right) c_{i,x}^2 = \Psi_{xx} \, \rho \, C\!\left(x + \frac{\Delta x}{2}\right),
\end{align}
with
\begin{align}
\label{form_C}
C = (\Delta t)^2 \left[ 2 T + (1-\theta) \frac{P_{\mathrm{EOS}} - \rho T}{\rho} + 2 \gamma_2 g^2 (\Delta t)^2 \right].
\end{align}
Figure~\ref{fig_ParamC} shows that $C(x)$ in Eq.~(\ref{form_C}) reasonably matches estimates from the second moments of $P(x,\delta x)$; error bars indicate percentage deviations from Fig.~\ref{fig_mom_compared}.  
In contrast, the same approach cannot directly extract $B$, since zeroth-order approximations for $\rho$ and $P$ distort $\sum_i f_i c_{i,x}$.  
Applying a first-order expansion of $\rho (\tilde{x})$ around $\rho(x)$ (for detailed formula, see Supplementary Material) reduces the maximum deviation in $\sum_i f_i c_{i,x}$ from 25\% to 3.3\%, showing that a first-order spatial correction for $\rho$ significantly improves the first-moment accuracy of $f_i$ in Eq.~(\ref{def_f_pbase}).

 %%%%%%%%%%%%%%%%%%%%%%%%%%%%%%%%%%%%%%%%%%%%%%%%%%%%

\section{Summary}
\label{summary}
\resubmitedit{
The consistency between molecular dynamics (MD) simulations and mesoscopic kinetic models for non-ideal fluids is investigated using the molecular dynamics lattice Boltzmann (MDLB) method.
In this approach, the discrete distribution function $f_i$ is directly constructed from MD simulations of particles confined between reflective walls under external forcing, enabling direct comparison with the LBM.
Our results indicate that moments of $f_i$ derived from MD are most closely reproduced by a mesoscopic model that blends pseudo-potential and free-energy formulations equally. 
Symbolic expressions for the first moments of the collision and force term, corresponding to spatial and temporal variations of $f_i$, and the second moment of $f_i$ (Eqs.~\ref{fst_omega_ansatz} and \ref{sec_f_ansatz}), match those of the hybrid formulation, reproducing macroscopic force balance at leading order. The optimal agreement is obtained at a blending ratio $\theta = 0.5$, suggesting that a balanced combination of the two approaches yields the most physically consistent mesoscopic representation.
Beyond leading order, analysis of Eq.~(\ref{sec_f_ansatz}) reveals higher-order corrections, including a $\rho g^2$ term. 
The best agreement occurs at $\gamma_3 = 0.25$, suggesting that the $S\!-\!C$ model effectively captures  microscopic behavior seen in MD simulations. This aligns with prior evidence that applying various EOS forms to the $S\!-\!C$ model yields thermodynamic behavior consistent with theory~\cite{Yuan_2006}. Alternatively, the EDM with a specific $\tau$ or a hybrid scheme combining approaches in Table~\ref{table_forceschemeparam} could also achieve $\gamma_3 = 0.25$.
The study further clarifies how microscopic statistical quantities contribute to macroscopic force balance. 
Specifically, the Gaussian probability distribution of particle displacements with mean \( B(x) \) and variance \( C(x) \) can accurately reproduce key moments of $f_i$ relevant to force balance. 
To achieve accuracy within a few percent, \( C(x) \)  must take the form given in Eq.~(\ref{form_C}), while \( B(x) \)  requires a first-order correction proportional to $\partial \rho / \partial x$.
The present results provide a direct connection between microscopic particle dynamics and mesoscopic modeling in non-ideal fluids. By identifying the force formulation that ensures consistency at the level of distribution function moments, this study offers a practical guideline for selecting and constructing LBM models. 
Building on this framework, future work will extend the analytic expressions developed here to systems undergoing phase separation, where higher-order contributions are expected to play a critical role in ensuring thermodynamic consistency and accurate mesoscopic modeling. Additional investigations will consider systems under background shear and non-isothermal conditions to further validate the framework in more general flow and thermal scenarios.
}

\section*{Acknowledgements}
H.O. sincerely acknowledges the support of Tufts University in providing access to its High-Performance Computing (HPC) Cluster, which enabled the simulations reported in this paper.

\nocite{*}
\bibliographystyle{apsrev4-2}
\bibliography{manuscript}% Produces the bibliography via BibTeX.

\providecommand{\noopsort}[1]{}\providecommand{\singleletter}[1]{#1}%
\begin{thebibliography}{30}%
\makeatletter
\providecommand \@ifxundefined [1]{%
 \@ifx{#1\undefined}
}%
\providecommand \@ifnum [1]{%
 \ifnum #1\expandafter \@firstoftwo
 \else \expandafter \@secondoftwo
 \fi
}%
\providecommand \@ifx [1]{%
 \ifx #1\expandafter \@firstoftwo
 \else \expandafter \@secondoftwo
 \fi
}%
\providecommand \natexlab [1]{#1}%
\providecommand \enquote  [1]{``#1''}%
\providecommand \bibnamefont  [1]{#1}%
\providecommand \bibfnamefont [1]{#1}%
\providecommand \citenamefont [1]{#1}%
\providecommand \href@noop [0]{\@secondoftwo}%
\providecommand \href [0]{\begingroup \@sanitize@url \@href}%
\providecommand \@href[1]{\@@startlink{#1}\@@href}%
\providecommand \@@href[1]{\endgroup#1\@@endlink}%
\providecommand \@sanitize@url [0]{\catcode `\\12\catcode `\$12\catcode
  `\&12\catcode `\#12\catcode `\^12\catcode `\_12\catcode `\%12\relax}%
\providecommand \@@startlink[1]{}%
\providecommand \@@endlink[0]{}%
\providecommand \url  [0]{\begingroup\@sanitize@url \@url }%
\providecommand \@url [1]{\endgroup\@href {#1}{\urlprefix }}%
\providecommand \urlprefix  [0]{URL }%
\providecommand \Eprint [0]{\href }%
\providecommand \doibase [0]{https://doi.org/}%
\providecommand \selectlanguage [0]{\@gobble}%
\providecommand \bibinfo  [0]{\@secondoftwo}%
\providecommand \bibfield  [0]{\@secondoftwo}%
\providecommand \translation [1]{[#1]}%
\providecommand \BibitemOpen [0]{}%
\providecommand \bibitemStop [0]{}%
\providecommand \bibitemNoStop [0]{.\EOS\space}%
\providecommand \EOS [0]{\spacefactor3000\relax}%
\providecommand \BibitemShut  [1]{\csname bibitem#1\endcsname}%
\let\auto@bib@innerbib\@empty
%</preamble>
\bibitem [{\citenamefont {Succi}(2018)}]{Succi2018}%
  \BibitemOpen
  \bibfield  {author} {\bibinfo {author} {\bibfnamefont {S.}~\bibnamefont
  {Succi}},\ }\href@noop {} {\emph {\bibinfo {title} {The Lattice Boltzmann
  Equation: For Complex States of Flowing Matter}}}\ (\bibinfo  {publisher}
  {Oxford University Press},\ \bibinfo {year} {2018})\BibitemShut {NoStop}%
\bibitem [{\citenamefont {Schiller}\ \emph {et~al.}(2018)\citenamefont
  {Schiller}, \citenamefont {Krüger},\ and\ \citenamefont
  {Henrich}}]{Schiller2018}%
  \BibitemOpen
  \bibfield  {author} {\bibinfo {author} {\bibfnamefont {U.~D.}\ \bibnamefont
  {Schiller}}, \bibinfo {author} {\bibfnamefont {T.}~\bibnamefont {Krüger}},\
  and\ \bibinfo {author} {\bibfnamefont {O.}~\bibnamefont {Henrich}},\ }\href
  {https://doi.org/10.1039/C7SM01711A} {\bibfield  {journal} {\bibinfo
  {journal} {Soft Matter}\ }\textbf {\bibinfo {volume} {14}},\ \bibinfo {pages}
  {9} (\bibinfo {year} {2018})}\BibitemShut {NoStop}%
\bibitem [{\citenamefont {Aidun}\ and\ \citenamefont
  {Clausen}(2010)}]{2010_review}%
  \BibitemOpen
  \bibfield  {author} {\bibinfo {author} {\bibfnamefont {C.~K.}\ \bibnamefont
  {Aidun}}\ and\ \bibinfo {author} {\bibfnamefont {J.~R.}\ \bibnamefont
  {Clausen}},\ }\href
  {https://doi.org/https://doi.org/10.1146/annurev-fluid-121108-145519}
  {\bibfield  {journal} {\bibinfo  {journal} {Annual Review of Fluid
  Mechanics}\ }\textbf {\bibinfo {volume} {42}},\ \bibinfo {pages} {439}
  (\bibinfo {year} {2010})}\BibitemShut {NoStop}%
\bibitem [{\citenamefont {Kr{\"u}ger}\ \emph {et~al.}(2017)\citenamefont
  {Kr{\"u}ger}, \citenamefont {Kusumaatmaja}, \citenamefont {Kuzmin},
  \citenamefont {Shardt}, \citenamefont {Silva},\ and\ \citenamefont
  {Viggen}}]{kruger2017lattice}%
  \BibitemOpen
  \bibfield  {author} {\bibinfo {author} {\bibfnamefont {T.}~\bibnamefont
  {Kr{\"u}ger}}, \bibinfo {author} {\bibfnamefont {H.}~\bibnamefont
  {Kusumaatmaja}}, \bibinfo {author} {\bibfnamefont {A.}~\bibnamefont
  {Kuzmin}}, \bibinfo {author} {\bibfnamefont {O.}~\bibnamefont {Shardt}},
  \bibinfo {author} {\bibfnamefont {G.}~\bibnamefont {Silva}},\ and\ \bibinfo
  {author} {\bibfnamefont {E.~M.}\ \bibnamefont {Viggen}},\ }\href
  {https://doi.org/10.1007/978-3-319-44649-3} {\emph {\bibinfo {title} {The
  Lattice Boltzmann Method: Principles and Practice}}}\ (\bibinfo  {publisher}
  {Springer},\ \bibinfo {address} {Cham, Switzerland},\ \bibinfo {year}
  {2017})\BibitemShut {NoStop}%
\bibitem [{\citenamefont {Otomo}(2023)}]{OTOMO2023}%
  \BibitemOpen
  \bibfield  {author} {\bibinfo {author} {\bibfnamefont {H.}~\bibnamefont
  {Otomo}},\ }\href@noop {} {\bibfield  {journal} {\bibinfo  {journal} {Journal
  of Statistical Physics}\ }\textbf {\bibinfo {volume} {190}},\ \bibinfo
  {pages} {112} (\bibinfo {year} {2023})}\BibitemShut {NoStop}%
\bibitem [{\citenamefont {Groot}\ and\ \citenamefont
  {Warren}(1997)}]{Groot_1997}%
  \BibitemOpen
  \bibfield  {author} {\bibinfo {author} {\bibfnamefont {R.~D.}\ \bibnamefont
  {Groot}}\ and\ \bibinfo {author} {\bibfnamefont {P.~B.}\ \bibnamefont
  {Warren}},\ }\href {https://doi.org/10.1063/1.474784} {\bibfield  {journal}
  {\bibinfo  {journal} {The Journal of Chemical Physics}\ }\textbf {\bibinfo
  {volume} {107}},\ \bibinfo {pages} {4423} (\bibinfo {year}
  {1997})}\BibitemShut {NoStop}%
\bibitem [{\citenamefont {Jamali}\ \emph {et~al.}(2015)\citenamefont {Jamali},
  \citenamefont {Boromand}, \citenamefont {Khani}, \citenamefont {Wagner},
  \citenamefont {Yamanoi},\ and\ \citenamefont {Maia}}]{Jamali_2015}%
  \BibitemOpen
  \bibfield  {author} {\bibinfo {author} {\bibfnamefont {S.}~\bibnamefont
  {Jamali}}, \bibinfo {author} {\bibfnamefont {A.}~\bibnamefont {Boromand}},
  \bibinfo {author} {\bibfnamefont {S.}~\bibnamefont {Khani}}, \bibinfo
  {author} {\bibfnamefont {J.}~\bibnamefont {Wagner}}, \bibinfo {author}
  {\bibfnamefont {M.}~\bibnamefont {Yamanoi}},\ and\ \bibinfo {author}
  {\bibfnamefont {J.}~\bibnamefont {Maia}},\ }\href
  {https://doi.org/10.1063/1.4919303} {\bibfield  {journal} {\bibinfo
  {journal} {The Journal of Chemical Physics}\ }\textbf {\bibinfo {volume}
  {142}},\ \bibinfo {pages} {164902} (\bibinfo {year} {2015})}\BibitemShut
  {NoStop}%
\bibitem [{\citenamefont {Zantop}\ and\ \citenamefont
  {Stark}(2021)}]{Zantop_2021}%
  \BibitemOpen
  \bibfield  {author} {\bibinfo {author} {\bibfnamefont {A.~W.}\ \bibnamefont
  {Zantop}}\ and\ \bibinfo {author} {\bibfnamefont {H.}~\bibnamefont {Stark}},\
  }\href {https://doi.org/10.1063/5.0064558} {\bibfield  {journal} {\bibinfo
  {journal} {The Journal of Chemical Physics}\ }\textbf {\bibinfo {volume}
  {155}},\ \bibinfo {pages} {134904} (\bibinfo {year} {2021})}\BibitemShut
  {NoStop}%
\bibitem [{\citenamefont {Denniston}\ and\ \citenamefont
  {Robbins}(2004)}]{Denniston_2004}%
  \BibitemOpen
  \bibfield  {author} {\bibinfo {author} {\bibfnamefont {C.}~\bibnamefont
  {Denniston}}\ and\ \bibinfo {author} {\bibfnamefont {M.~O.}\ \bibnamefont
  {Robbins}},\ }\href {https://doi.org/10.1103/PhysRevE.69.021505} {\bibfield
  {journal} {\bibinfo  {journal} {Phys. Rev. E}\ }\textbf {\bibinfo {volume}
  {69}},\ \bibinfo {pages} {021505} (\bibinfo {year} {2004})}\BibitemShut
  {NoStop}%
\bibitem [{\citenamefont {Tong}\ \emph {et~al.}(2022)\citenamefont {Tong},
  \citenamefont {Li},\ and\ \citenamefont {Li}}]{Tong_2022}%
  \BibitemOpen
  \bibfield  {author} {\bibinfo {author} {\bibfnamefont {Z.}~\bibnamefont
  {Tong}}, \bibinfo {author} {\bibfnamefont {M.}~\bibnamefont {Li}},\ and\
  \bibinfo {author} {\bibfnamefont {D.}~\bibnamefont {Li}},\ }\href
  {https://doi.org/10.1615/HEATTRANSRES.2021041645} {\bibfield  {journal}
  {\bibinfo  {journal} {Heat Transfer Research}\ }\textbf {\bibinfo {volume}
  {53}},\ \bibinfo {pages} {33} (\bibinfo {year} {2022})}\BibitemShut {NoStop}%
\bibitem [{\citenamefont {Gunstensen}\ \emph {et~al.}(1991)\citenamefont
  {Gunstensen}, \citenamefont {Rothman}, \citenamefont {Zaleski},\ and\
  \citenamefont {Zanetti}}]{Gunstensen_1991}%
  \BibitemOpen
  \bibfield  {author} {\bibinfo {author} {\bibfnamefont {A.~K.}\ \bibnamefont
  {Gunstensen}}, \bibinfo {author} {\bibfnamefont {D.~H.}\ \bibnamefont
  {Rothman}}, \bibinfo {author} {\bibfnamefont {S.}~\bibnamefont {Zaleski}},\
  and\ \bibinfo {author} {\bibfnamefont {G.}~\bibnamefont {Zanetti}},\ }\href
  {https://doi.org/10.1103/PhysRevA.43.4320} {\bibfield  {journal} {\bibinfo
  {journal} {Phys. Rev. A}\ }\textbf {\bibinfo {volume} {43}},\ \bibinfo
  {pages} {4320} (\bibinfo {year} {1991})}\BibitemShut {NoStop}%
\bibitem [{\citenamefont {Shan}\ and\ \citenamefont
  {Chen}(1993)}]{Shan-Chen_1993}%
  \BibitemOpen
  \bibfield  {author} {\bibinfo {author} {\bibfnamefont {X.}~\bibnamefont
  {Shan}}\ and\ \bibinfo {author} {\bibfnamefont {H.}~\bibnamefont {Chen}},\
  }\href {https://doi.org/10.1103/PhysRevE.47.1815} {\bibfield  {journal}
  {\bibinfo  {journal} {Phys. Rev. E}\ }\textbf {\bibinfo {volume} {47}},\
  \bibinfo {pages} {1815} (\bibinfo {year} {1993})}\BibitemShut {NoStop}%
\bibitem [{\citenamefont {Swift}\ \emph {et~al.}(1995)\citenamefont {Swift},
  \citenamefont {Osborn},\ and\ \citenamefont {Yeomans}}]{Swift_1995}%
  \BibitemOpen
  \bibfield  {author} {\bibinfo {author} {\bibfnamefont {M.~R.}\ \bibnamefont
  {Swift}}, \bibinfo {author} {\bibfnamefont {W.~R.}\ \bibnamefont {Osborn}},\
  and\ \bibinfo {author} {\bibfnamefont {J.~M.}\ \bibnamefont {Yeomans}},\
  }\href {https://doi.org/10.1103/PhysRevLett.75.830} {\bibfield  {journal}
  {\bibinfo  {journal} {Phys. Rev. Lett.}\ }\textbf {\bibinfo {volume} {75}},\
  \bibinfo {pages} {830} (\bibinfo {year} {1995})}\BibitemShut {NoStop}%
\bibitem [{\citenamefont {Swift}\ \emph {et~al.}(1996)\citenamefont {Swift},
  \citenamefont {Orlandini}, \citenamefont {Osborn},\ and\ \citenamefont
  {Yeomans}}]{Swift_1996}%
  \BibitemOpen
  \bibfield  {author} {\bibinfo {author} {\bibfnamefont {M.~R.}\ \bibnamefont
  {Swift}}, \bibinfo {author} {\bibfnamefont {E.}~\bibnamefont {Orlandini}},
  \bibinfo {author} {\bibfnamefont {W.~R.}\ \bibnamefont {Osborn}},\ and\
  \bibinfo {author} {\bibfnamefont {J.~M.}\ \bibnamefont {Yeomans}},\ }\href
  {https://doi.org/10.1103/PhysRevE.54.5041} {\bibfield  {journal} {\bibinfo
  {journal} {Phys. Rev. E}\ }\textbf {\bibinfo {volume} {54}},\ \bibinfo
  {pages} {5041} (\bibinfo {year} {1996})}\BibitemShut {NoStop}%
\bibitem [{\citenamefont {Guo}\ \emph {et~al.}(2002)\citenamefont {Guo},
  \citenamefont {Zheng},\ and\ \citenamefont {Shi}}]{Guo_2002}%
  \BibitemOpen
  \bibfield  {author} {\bibinfo {author} {\bibfnamefont {Z.}~\bibnamefont
  {Guo}}, \bibinfo {author} {\bibfnamefont {C.}~\bibnamefont {Zheng}},\ and\
  \bibinfo {author} {\bibfnamefont {B.}~\bibnamefont {Shi}},\ }\href
  {https://doi.org/10.1103/PhysRevE.65.046308} {\bibfield  {journal} {\bibinfo
  {journal} {Phys. Rev. E}\ }\textbf {\bibinfo {volume} {65}},\ \bibinfo
  {pages} {046308} (\bibinfo {year} {2002})}\BibitemShut {NoStop}%
\bibitem [{\citenamefont {Kupershtokh}\ \emph {et~al.}(2009)\citenamefont
  {Kupershtokh}, \citenamefont {Medvedev},\ and\ \citenamefont
  {Karpov}}]{EDM_2009}%
  \BibitemOpen
  \bibfield  {author} {\bibinfo {author} {\bibfnamefont {A.}~\bibnamefont
  {Kupershtokh}}, \bibinfo {author} {\bibfnamefont {D.}~\bibnamefont
  {Medvedev}},\ and\ \bibinfo {author} {\bibfnamefont {D.}~\bibnamefont
  {Karpov}},\ }\href
  {https://doi.org/https://doi.org/10.1016/j.camwa.2009.02.024} {\bibfield
  {journal} {\bibinfo  {journal} {Computers and Mathematics with Applications}\
  }\textbf {\bibinfo {volume} {58}},\ \bibinfo {pages} {965} (\bibinfo {year}
  {2009})},\ \bibinfo {note} {mesoscopic Methods in Engineering and
  Science}\BibitemShut {NoStop}%
\bibitem [{\citenamefont {Parsa}\ and\ \citenamefont
  {Wagner}(2017)}]{Alex_2017}%
  \BibitemOpen
  \bibfield  {author} {\bibinfo {author} {\bibfnamefont {M.~R.}\ \bibnamefont
  {Parsa}}\ and\ \bibinfo {author} {\bibfnamefont {A.~J.}\ \bibnamefont
  {Wagner}},\ }\href {https://doi.org/10.1103/PhysRevE.96.013314} {\bibfield
  {journal} {\bibinfo  {journal} {Phys. Rev. E}\ }\textbf {\bibinfo {volume}
  {96}},\ \bibinfo {pages} {013314} (\bibinfo {year} {2017})}\BibitemShut
  {NoStop}%
\bibitem [{\citenamefont {Thompson}\ \emph {et~al.}(2022)\citenamefont
  {Thompson}, \citenamefont {Aktulga}, \citenamefont {Berger}, \citenamefont
  {Bolintineanu}, \citenamefont {Brown}, \citenamefont {Crozier}, \citenamefont
  {in~'t Veld}, \citenamefont {Kohlmeyer}, \citenamefont {Moore}, \citenamefont
  {Nguyen}, \citenamefont {Shan}, \citenamefont {Stevens}, \citenamefont
  {Tranchida}, \citenamefont {Trott},\ and\ \citenamefont {Plimpton}}]{LAMMPS}%
  \BibitemOpen
  \bibfield  {author} {\bibinfo {author} {\bibfnamefont {A.~P.}\ \bibnamefont
  {Thompson}}, \bibinfo {author} {\bibfnamefont {H.~M.}\ \bibnamefont
  {Aktulga}}, \bibinfo {author} {\bibfnamefont {R.}~\bibnamefont {Berger}},
  \bibinfo {author} {\bibfnamefont {D.~S.}\ \bibnamefont {Bolintineanu}},
  \bibinfo {author} {\bibfnamefont {W.~M.}\ \bibnamefont {Brown}}, \bibinfo
  {author} {\bibfnamefont {P.~S.}\ \bibnamefont {Crozier}}, \bibinfo {author}
  {\bibfnamefont {P.~J.}\ \bibnamefont {in~'t Veld}}, \bibinfo {author}
  {\bibfnamefont {A.}~\bibnamefont {Kohlmeyer}}, \bibinfo {author}
  {\bibfnamefont {S.~G.}\ \bibnamefont {Moore}}, \bibinfo {author}
  {\bibfnamefont {T.~D.}\ \bibnamefont {Nguyen}}, \bibinfo {author}
  {\bibfnamefont {R.}~\bibnamefont {Shan}}, \bibinfo {author} {\bibfnamefont
  {M.~J.}\ \bibnamefont {Stevens}}, \bibinfo {author} {\bibfnamefont
  {J.}~\bibnamefont {Tranchida}}, \bibinfo {author} {\bibfnamefont
  {C.}~\bibnamefont {Trott}},\ and\ \bibinfo {author} {\bibfnamefont {S.~J.}\
  \bibnamefont {Plimpton}},\ }\href {https://doi.org/10.1016/j.cpc.2021.108171}
  {\bibfield  {journal} {\bibinfo  {journal} {Comp. Phys. Comm.}\ }\textbf
  {\bibinfo {volume} {271}},\ \bibinfo {pages} {108171} (\bibinfo {year}
  {2022})}\BibitemShut {NoStop}%
\bibitem [{\citenamefont {Plimpton}(1995)}]{plimpton1995fast}%
  \BibitemOpen
  \bibfield  {author} {\bibinfo {author} {\bibfnamefont {S.}~\bibnamefont
  {Plimpton}},\ }\href@noop {} {\bibfield  {journal} {\bibinfo  {journal}
  {Journal of Computational Physics}\ }\textbf {\bibinfo {volume} {117}},\
  \bibinfo {pages} {1} (\bibinfo {year} {1995})}\BibitemShut {NoStop}%
\bibitem [{\citenamefont {Sun}\ and\ \citenamefont {Teja}(1996)}]{Teja_1996}%
  \BibitemOpen
  \bibfield  {author} {\bibinfo {author} {\bibfnamefont {T.}~\bibnamefont
  {Sun}}\ and\ \bibinfo {author} {\bibfnamefont {A.~S.}\ \bibnamefont {Teja}},\
  }\href@noop {} {\bibfield  {journal} {\bibinfo  {journal} {J. Phys. Chem.}\
  }\textbf {\bibinfo {volume} {100}},\ \bibinfo {pages} {17365} (\bibinfo
  {year} {1996})}\BibitemShut {NoStop}%
\bibitem [{\citenamefont {Otomo}\ \emph {et~al.}(2017)\citenamefont {Otomo},
  \citenamefont {Boghosian},\ and\ \citenamefont {Dubois}}]{OTOMO20171000}%
  \BibitemOpen
  \bibfield  {author} {\bibinfo {author} {\bibfnamefont {H.}~\bibnamefont
  {Otomo}}, \bibinfo {author} {\bibfnamefont {B.~M.}\ \bibnamefont
  {Boghosian}},\ and\ \bibinfo {author} {\bibfnamefont {F.}~\bibnamefont
  {Dubois}},\ }\href
  {https://doi.org/https://doi.org/10.1016/j.physa.2017.06.010} {\bibfield
  {journal} {\bibinfo  {journal} {Physica A: Statistical Mechanics and its
  Applications}\ }\textbf {\bibinfo {volume} {486}},\ \bibinfo {pages} {1000}
  (\bibinfo {year} {2017})}\BibitemShut {NoStop}%
\bibitem [{\citenamefont {Wagner}(2006)}]{Alex_2006}%
  \BibitemOpen
  \bibfield  {author} {\bibinfo {author} {\bibfnamefont {A.~J.}\ \bibnamefont
  {Wagner}},\ }\href {https://doi.org/10.1103/PhysRevE.74.056703} {\bibfield
  {journal} {\bibinfo  {journal} {Phys. Rev. E}\ }\textbf {\bibinfo {volume}
  {74}},\ \bibinfo {pages} {056703} (\bibinfo {year} {2006})}\BibitemShut
  {NoStop}%
\bibitem [{\citenamefont {Shan}(2008)}]{Shan_2008}%
  \BibitemOpen
  \bibfield  {author} {\bibinfo {author} {\bibfnamefont {X.}~\bibnamefont
  {Shan}},\ }\href {https://doi.org/10.1103/PhysRevE.77.066702} {\bibfield
  {journal} {\bibinfo  {journal} {Phys. Rev. E}\ }\textbf {\bibinfo {volume}
  {77}},\ \bibinfo {pages} {066702} (\bibinfo {year} {2008})}\BibitemShut
  {NoStop}%
\bibitem [{\citenamefont {Khajepor}\ \emph {et~al.}(2015)\citenamefont
  {Khajepor}, \citenamefont {Wen},\ and\ \citenamefont {Chen}}]{Baixin_2015}%
  \BibitemOpen
  \bibfield  {author} {\bibinfo {author} {\bibfnamefont {S.}~\bibnamefont
  {Khajepor}}, \bibinfo {author} {\bibfnamefont {J.}~\bibnamefont {Wen}},\ and\
  \bibinfo {author} {\bibfnamefont {B.}~\bibnamefont {Chen}},\ }\href
  {https://doi.org/10.1103/PhysRevE.91.023301} {\bibfield  {journal} {\bibinfo
  {journal} {Phys. Rev. E}\ }\textbf {\bibinfo {volume} {91}},\ \bibinfo
  {pages} {023301} (\bibinfo {year} {2015})}\BibitemShut {NoStop}%
\bibitem [{\citenamefont {Parsa}\ \emph {et~al.}(2019)\citenamefont {Parsa},
  \citenamefont {Pachalieva},\ and\ \citenamefont {Wagner}}]{Alex_2019}%
  \BibitemOpen
  \bibfield  {author} {\bibinfo {author} {\bibfnamefont {M.~R.}\ \bibnamefont
  {Parsa}}, \bibinfo {author} {\bibfnamefont {A.}~\bibnamefont {Pachalieva}},\
  and\ \bibinfo {author} {\bibfnamefont {A.~J.}\ \bibnamefont {Wagner}},\
  }\href {https://doi.org/10.1142/S0129183119410079} {\bibfield  {journal}
  {\bibinfo  {journal} {International Journal of Modern Physics C}\ }\textbf
  {\bibinfo {volume} {30}},\ \bibinfo {pages} {1941007} (\bibinfo {year}
  {2019})}\BibitemShut {NoStop}%
\bibitem [{\citenamefont {Pachalieva}\ and\ \citenamefont
  {Wagner}(2020)}]{Alex_2020_PRE}%
  \BibitemOpen
  \bibfield  {author} {\bibinfo {author} {\bibfnamefont {A.}~\bibnamefont
  {Pachalieva}}\ and\ \bibinfo {author} {\bibfnamefont {A.~J.}\ \bibnamefont
  {Wagner}},\ }\href {https://doi.org/10.1103/PhysRevE.102.053310} {\bibfield
  {journal} {\bibinfo  {journal} {Phys. Rev. E}\ }\textbf {\bibinfo {volume}
  {102}},\ \bibinfo {pages} {053310} (\bibinfo {year} {2020})}\BibitemShut
  {NoStop}%
\bibitem [{\citenamefont {Parsa}\ and\ \citenamefont
  {Wagner}(2020)}]{Alex_2020}%
  \BibitemOpen
  \bibfield  {author} {\bibinfo {author} {\bibfnamefont {M.~R.}\ \bibnamefont
  {Parsa}}\ and\ \bibinfo {author} {\bibfnamefont {A.~J.}\ \bibnamefont
  {Wagner}},\ }\href {https://doi.org/10.1103/PhysRevLett.124.234501}
  {\bibfield  {journal} {\bibinfo  {journal} {Phys. Rev. Lett.}\ }\textbf
  {\bibinfo {volume} {124}},\ \bibinfo {pages} {234501} (\bibinfo {year}
  {2020})}\BibitemShut {NoStop}%
\bibitem [{\citenamefont {Pachalieva}\ and\ \citenamefont
  {Wagner}(2021{\natexlab{a}})}]{Alex_2021}%
  \BibitemOpen
  \bibfield  {author} {\bibinfo {author} {\bibfnamefont {A.}~\bibnamefont
  {Pachalieva}}\ and\ \bibinfo {author} {\bibfnamefont {A.~J.}\ \bibnamefont
  {Wagner}},\ }\href {https://arxiv.org/abs/2109.05009} {\bibinfo {title}
  {Connecting lattice boltzmann methods to physical reality by coarse-graining
  molecular dynamics simulations}} (\bibinfo {year} {2021}{\natexlab{a}}),\
  \Eprint {https://arxiv.org/abs/2109.05009} {arXiv:2109.05009
  [physics.comp-ph]} \BibitemShut {NoStop}%
\bibitem [{\citenamefont {Pachalieva}\ and\ \citenamefont
  {Wagner}(2021{\natexlab{b}})}]{Alex_2021_Phil}%
  \BibitemOpen
  \bibfield  {author} {\bibinfo {author} {\bibfnamefont {A.}~\bibnamefont
  {Pachalieva}}\ and\ \bibinfo {author} {\bibfnamefont {A.~J.}\ \bibnamefont
  {Wagner}},\ }\href {https://doi.org/10.1098/rsta.2020.0404} {\bibfield
  {journal} {\bibinfo  {journal} {Philosophical Transactions of the Royal
  Society A: Mathematical, Physical and Engineering Sciences}\ }\textbf
  {\bibinfo {volume} {379}},\ \bibinfo {pages} {20200404} (\bibinfo {year}
  {2021}{\natexlab{b}})}\BibitemShut {NoStop}%
\bibitem [{\citenamefont {Yuan}\ and\ \citenamefont
  {Schaefer}(2006)}]{Yuan_2006}%
  \BibitemOpen
  \bibfield  {author} {\bibinfo {author} {\bibfnamefont {P.}~\bibnamefont
  {Yuan}}\ and\ \bibinfo {author} {\bibfnamefont {L.}~\bibnamefont
  {Schaefer}},\ }\href {https://doi.org/10.1063/1.2187070} {\bibfield
  {journal} {\bibinfo  {journal} {Physics of Fluids}\ }\textbf {\bibinfo
  {volume} {18}},\ \bibinfo {pages} {042101} (\bibinfo {year} {2006})},\
  \Eprint
  {https://arxiv.org/abs/https://pubs.aip.org/aip/pof/article-pdf/doi/10.1063/1.2187070/13662193/042101\_1\_online.pdf}
  {https://pubs.aip.org/aip/pof/article-pdf/doi/10.1063/1.2187070/13662193/042101\_1\_online.pdf}
  \BibitemShut {NoStop}%
\bibitem{supplemental}
See Supplemental Material at \url{[URL]} for additional simulation data, including $a^2$ profiles, comparisons of distribution-function moments with theoretical predictions under different forcing and numerical conditions, and analytical derivations of $\Psi_{xx}$ and the mean displacement $B(x)$.
\end{thebibliography}%

\end{document}